\documentclass[prb,aps,twocolumn,floats]{revtex4}
\usepackage{amstext}
\usepackage{amsmath}
\usepackage{graphics}
\usepackage{multicol}
\usepackage{epsfig}

\begin{document}

\title{Elasticity of nanometer-sized objects}

\author{D.E. Segall} \address{Department of Physics, Massachusetts Institute of Technology,
Cambridge MA 02139}

\author{Sohrab Ismail-Beigi} \address{Department of Physics, University of California at Berkeley,
Berkeley, CA 94720}

\author{T.A. Arias}  \address{Laboratory of Atomic and Solid State Physics, Cornell University,
Ithaca, NY 14853}

\begin{abstract}
We initiate the development of a theory of the elasticity of nanoscale
objects based upon new physical concepts which remain properly defined
on the nanoscale.  This theory provides a powerful way of
understanding nanoscale elasticity in terms of local group
contributions and gives insight into the breakdown of standard
continuum relations.  We also give two applications.  In the first, we
show how to use the theory to derive a new relation between the
bending and stretching properties of nanomechanical resonators and to
prove that it is much more accurate than the continuum-based relations
currently employed in present experimental analyses.  In the second,
we use the new approach to link features of the underlining electronic
structure to the elastic response of a silicon nanoresonator.
\end{abstract}

\twocolumn

\maketitle

\def\ctbt{{$c(2\times 2)$ }}
\def\tbo{$2\times 1$ }

\section{Introduction} \label{sec-intro}
The recent development of artificial free-standing structures of
nanometer dimensions has led to great interest in their mechanical
properties.  A wealth of experimental information is now available for
nanowires \cite{wong,roukes,osakabe,craighead} and
nanotubes\cite{wong,treacy,salvetat}, and a computational literature
is developing on the
subject\cite{yakobson,hernandez,lu,robertson,mon_a,mon_b,broughton}.
Many of these works make use of results from the continuum theory of
elasticity to analyze the behavior of nanometer structures.  However,
the applicability of continuum theories to nanoscale objects, where
atomic-level inhomogeneities come to the fore, has yet to be explored
in depth.

Rigorous understanding of the elastic properties of nanoscale systems
is crucial in understanding their mechanical behavior and presents an
intriguing theoretical challenge lying at the cross-over between the
atomic level and the continuum.  In the absence of an appropriate
theoretical description at this cross-over, critical questions remain
to be answered including the extent to which continuum theories can be
pushed into the nanoregime, how to provide systematic corrections to
continuum theory, what effects do different bonding arrangements have
on elastic response, and what signatures in the electronic structure
correlate with the mechanical properties of the overall structure?

Recently, there have been a number of theoretical explorations of the
impact of nanoscale structure on mechanical
properties\cite{broughton,miller,vitek_a,vitek_b,vitek_c,jayanthi}.
These studies fall under two broad approaches, either the addition of
surface and edge corrections to bulk continuum
theories\cite{broughton,miller} or the extraction of overall
mechanical response from atomic scale
interactions\cite{vitek_a,vitek_b,vitek_c,jayanthi}.  The latter
approach has the distinct advantage of allowing first principles
understanding of how different chemical groups and bonding
arrangements contribute to overall elastic response, thus opening the
potential for the rational design of nanostructures with specific
properties.

In coarse graining from interatomic interactions to mechanical
response, some works rely upon the problematic decomposition of the
total system energy into a direct sum of atomic
energies\cite{vitek_a,vitek_b}, which is always arbitrary and
particularly inconvenient for connection with {\em ab initio}
electronic structure calculations.  The remaining works which attempt
to build up overall response from atomic level
contributions\cite{vitek_c,jayanthi} fail to account properly for the
Poisson effect.  Below we show that failure to account for this effect
leads to surprisingly unphysical results.

This manuscript presents the first theory for the analysis of overall
mechanical response in terms of atomic-level observables which suffers
from neither difficulty from the preceding paragraph.  This analysis
allows, for the first time, quantitative understanding of how
continuum theory breaks down on the nanoscale, of how to make
appropriate corrections, and of how to predict the effects of
different bonding arrangements on overall elastic response.  It is
well known that the decomposition of overall elastic response into a
sum of atomic level contributions is not unique.  We show here,
however, that with the additional constraint of dependence of moduli
on local environment our definition of atomic level moduli becomes
physically meaningful and essentially unique when coarse-grained over
regions of extent comparable to the decay range of the force-constant
matrix.

For concreteness, in this work we focus on nanowires.  However, we
will also describe briefly how this work can be extended to any system
with nanometer dimensions. The manuscript proceeds as follows.
Section~\ref{overview} briefly overviews the present state of the
field.  Next, as the traditional concept of Young's modulus becomes
ill-defined on the nanoscale, we begin by carefully defining continuum
elastic constants appropriate for nanowires in Section~\ref{sec-nr}.
We then show how to decompose these constants exactly into
atomic-level contributions based on true physical observables (rather
than individual atomic energies) using a straight-forward application
of Born and Huang's method of long waves~\cite{born}, resulting in a
decomposition similar in spirit to those in References~[17]~and~[18]
(Section~\ref{sec-mlw}).  Section~\ref{sec-fail} demonstrates the
surprising, radical breakdown of this approach when applied to
nanoresonators.  Then, in Section~\ref{sec-mlwn}, we identify the
source of the difficulty as the Poisson effect and present the first
analysis of mechanical response truly applicable to nanoresonators.

The manuscript then goes on to applications.  Section~\ref{sec-trans}
verifies the physical meaningfulness of our newly defined quantities
by verifying that they predict response to modes of strain for which
they were not directly constructed.  We then, in
Section~\ref{sec-fte}, use our approach to generate a new, much more
accurate, relationship between experimentally accessible observables
describing response to flexural and extensional strain in
nanomechanical resonators.  Finally, Section~\ref{sec-cbames} uses
this theory to explore possible links between underlying electronic
structure and local elastic response.

\section{Overview} \label{overview}

As the introduction mentions, the literature pursues two broad
categories of approach to the study of mechanical properties on the
nanoscale, either surface and edge corrections to continuum theory or
extraction of overall response from the underlying atomic
interactions.  In the former category, Reference [13], through scaling
arguments and numerical examples, notes that the Young's modulus for
nanomechanical resonators scales as a bulk term plus surface and edge
corrections.  Although providing insight and motivation, this work
leaves completely open how one should understand these corrections
from first principles.  Reference [14] provides a more rigorous study
based on separating nanoscale systems into continuum surface and bulk
regions.  This latter approach allows prediction of changes in
stiffness properties as one approaches nanometer length scales and has
the appeal of generating physically motivated correction terms.
However, it relies on the separation of a nanomechanical resonator
into bulk and surface continua as an {\em ansatz} and therefore
neither predicts when such a picture suffices to give an accurate
description nor prescribes further corrections.  

References [15-18], on the other hand, start from the more general
atomic level description and then try to unveil physical properties
from the underlying atomic description.  It is important to note that
these works do not deal directly with nanoscale systems but rather
focus on the effects of nanoscale inhomogeneities in {\em bulk}
systems.

References [15] and [16] concern the elastic properties of grain
boundaries.  These works define atomic-level elastic moduli as the
second derivative of the energy associated with each atom with respect
to strain and then go on to study the behavior of such moduli near
grain boundaries.  The difficulty with this approach is that it
requires a breakdown into individual contributions from each atom of
the total energy of any system.  Such an atomic energy is neither
observable nor uniquely defined and therefore cannot serve as an
appropriate basis for theoretical understanding.

Although References [17] and [18] work from valid physical
observables, the components of the force-constant matrix, these works
focus on bulk-like or mesoscopic scale systems and fail for nanoscale
systems for the reasons which we describe in this work.  Reference
[17] investigates nonlocal elastic constants on the mesoscopic scale
and links them to the underlying atomic interactions.  It then
proceeds to define an elastic constant for each atom and studies the
behavior of these quantities near surfaces and grain boundaries.
Reference [18] defines a bond frequency from the force-constant matrix
from which it deduces the possibility of bond rupture during crack
nucleation.  Neither of the above works properly accounts for the
Poisson effect, which we show in
Sections~\ref{sec-mlw},~\ref{sec-fail}~and~\ref{sec-mlwn} to play a
critical role in the elasticity of nanoscale systems.  Moreover,
straightforward generalization of these works to include this effect
fail for the the same reasons as does the related approach which we
describe in Section~\ref{sec-mlw}.

\section{Nanowire rigidities}  \label{sec-nr}
The prime difficulty in the application of continuum theory to objects
of nanometer cross-section is the loss of the ratio of the
inter-atomic spacing to the cross-sectional dimension as a small
parameter.  However, so long as the length of an object and the
wavelength of the distortions considered both greatly exceed the
inter-atomic spacing and the cross-sectional dimension, the object
properly may be viewed as a one-dimensional continuum.  Although we
focus in this work on nanowires, the generalization of the discussion
below to nanoscale systems of other dimensionality such as thin plates
or nanoscopic objects is straightforward.

Viewed as a linear continuum, the free energy per unit length $f$ of a
nanowire is
\begin{equation}
f = (E u^2 + F R^{-2}+T \tau^2)/2, \label{eqn:free}
\end{equation}
where $u$ is the linear strain of extension, $R$ is the radius of
curvature and $\tau$ is the rate of twist of the torsion.  The
coupling constant $E$ is the {\em extensional rigidity}, $F$ is the
{\em flexural rigidity} and $T$ is the {\em torsional rigidity}.
Unlike traditional bulk continuum concepts, the free-energy function
Eq.~(\ref{eqn:free}) is observable in principle and thus provides an
unambiguous operational definition of the rigidities.  We avoid the
use of traditional continuum concepts, such as the Young's modulus and
the cross-sectional area, because such concepts are neither uniquely
nor well-defined for nanoscale systems.

The rigidities in Eq.~(\ref{eqn:free}) are related to the phonon
frequencies through
\begin{eqnarray} 
\omega_{LA} & = & \sqrt{E/(\lambda m)} q, \label{eqn:mlw} \\
\omega_{TA} & = &\sqrt{F/(\lambda m)} q^2,  \label{eqn:mtw} \\
\omega_{RA} & = &\sqrt{T/(\lambda m I_r)} q, \label{eqn:mrw}  
\end{eqnarray}
where $\omega$ is the frequency for either the longitudinal,
transverse or rotational acoustic modes, respectively, $\lambda$ is
the linear atomic number density, $q$ is the wave vector and $m$ is
the mass of a single atom.  (This work focuses on single species
systems for simplicity.)  Finally, $I_r$ is defined unambiguously as
the mean rotational moment $I_r = (1/N_c) \sum_{\alpha} (x^2_{\alpha}
+ y^2_{\alpha})$, where the sum ranges over all atoms in the cell,
$N_c$ is the number of atoms per unit cell, the wire is assumed to run
along the $z$-axis and the origin lies on the center line of the wire.

Finally we note that although the rigidities in Eq.~(\ref{eqn:free})
are well-defined, certain traditional continuum relations between them
do not hold.  Specifically, we will show below that the traditional
continuum relationship between the extensional rigidity $E$ and the
flexural rigidity $F$ fails on the nanoscale, similar to relations
which have been recently used in the analysis of
experiments\cite{wong,treacy,salvetat,osakabe}.

\section{Method of long waves} \label{sec-mlw}
One reason for breakdown of traditional continuum relations on the
nanoscale is that the continuum perspective course grains away
important fluctuations which occur over distances on the order of the
inter-atomic spacing.  To overcome this shortcoming, we propose to
coarse grain only on distances over which the underlying interatomic
interactions vary, the decay length of the force-constant matrix.  The
straightforward approach to generate such a theory is the ``method of
long waves'' developed by Born and Huang\cite{born}, which is somewhat
similar to the approaches which have been used previously to defects
in bulk systems\cite{vitek_c,jayanthi}.  This section
applies the method of long waves to nanoresonators.  The next section
shows how, surprisingly, this and related approaches fail in the study
of systems with free surfaces and therefore, in general, cannot be
used to describe nanoscale systems.  In Section~\ref{sec-mlwn}, we
describe how to go beyond the straightforward application of the
``method of long waves'' in order to achieve a meaningful description.

We focus initially on longitudinal waves and, as noted in
Section~\ref{sec-nr}, choose our origin to lie on the center line of
the wire, which we let run along the $z$-axis.  For nanowires, the
presence of surfaces breaks periodicity in the transverse directions
resulting in a one-dimensional crystal with an extremely large unit
cell of length $L_c=N_c/\lambda$, where $N_c$ and $\lambda$ are as in
Section~\ref{sec-nr}.  In all expressions below, boldfaced quantities
are $3N_c$-dimensional and arrowed vector quantities are
three-dimensional.  Finally, sums with Greek indices range over atoms
in the unit cell.

To relate the rigidities to the dynamical matrix, we begin similarly
to Born and Huang and choose to factor the Bloch phases ($e^{iqz}$)
out of the representation of the phonon polarization vector ${\bf u}$,
incorporating them into the definition of the dynamical matrix ${\bf
D}$, so that the acoustic phonon polarization vectors are periodic
across the cell boundaries.  This ensures a uniform description of the
distribution of elastic energy along the axis of the wire.  To
generate a scalar equation for the phonon frequency, Born and Huang
project the secular equation for the dynamical matrix,
\begin{equation}
{\bf D} {\bf u}=-m \omega^2 {\bf u}, \label{eqn:sec}
\end{equation}
against the zeroth-order polarization vector ${\bf u}^{[0]}$.  Here,
however, to more symmetrically represent the distribution of elastic
energy, we project against the full polarization vector ${\bf u}$. 
Equating the frequency $\omega$ in Eq.~(\ref{eqn:sec}) with the
longitudinal frequency in Eq.~(\ref{eqn:mlw}) gives
\begin{equation}
-\frac{E}{\lambda}  =  \frac{[{\bf u}^\dagger {\bf D} {\bf u}]^{[2]}}{{\bf u}^\dagger {\bf
u}}  =\sum_{s,t=0}^{1} \frac{(-1)^{s}}{N_c} \: {\bf
u}^{\dagger[s]} \: {\bf
D}^{[2-s-t]} \: {\bf u}^{[t]}, \label{eqn:omegasq} 
\end{equation}
where we have expanded the numerator of the Rayleigh quotient to
second-order in powers of $(iq)$ and where the $3\times 3$ sub-block
of ${\bf D}^{[n]}$, which couples atoms $\alpha$ and $\beta$, is
\begin{equation} \label{eqn:Dn}
[{\bf D}^{[n]}]_{\alpha\beta}=\frac{1}{n!} \sum_{\vec R}{
\Phi_{\alpha\beta}(\vec R) \left({\hat z} \cdot(\vec R + \vec \tau_\beta - \vec \tau_\alpha)
\right)^n}.
\end{equation}
Here, $\vec R$ is a lattice vector along the $z-axis$,
$\Phi_{\alpha\beta}(\vec R)$ is the $3\times 3$ sub-block of the
force-constant matrix which couples atoms $\alpha$ and $\beta$ located
at positions $\vec \tau_{\alpha}$ and $\vec R + \vec \tau_{\beta}$,
respectively, and
\begin{equation}
\Phi_{\alpha\beta}(\vec R) = -\frac{\partial^2 U}{\partial \vec
\tau_{\alpha} \partial (\vec R + \vec \tau_{\beta})}. \nonumber
\end{equation} 

Finally, substituting Eq.~(\ref{eqn:Dn}) and $[{\bf
u}^{[0]}]_\alpha=\hat z$ into Eq.~(\ref{eqn:omegasq}), allows us to
express $E$ as a sum over atoms ($\alpha$) in the unit cell and all
atoms ($\beta,\vec R$) in the system,
\begin{eqnarray} 
E & = & \frac{1}{L_c} \sum_\alpha e_\alpha \label{eqn:thermofree} \\ 
e_\alpha & = &
\sum_{\beta \vec{R}} \mbox{\Large\{} - {{\Delta \vec Z}_{\alpha\beta}} \cdot
\Phi_{\alpha\beta}(\vec{R})\cdot {{\Delta \vec Z}_{\alpha\beta}} /2 + {\vec u^{[1]}_\alpha} \cdot
\Phi_{\alpha\beta}(\vec{R})\cdot {\vec u^{[1]}_\beta} \nonumber\\ && + {{\Delta \vec Z}_{\alpha\beta}}
\cdot \Phi_{\alpha\beta}(\vec{R})\cdot {\vec u^{[1]}_\beta} - {\vec u^{[1]}_\alpha} \cdot
\Phi_{\alpha\beta}(\vec{R})\cdot {{\Delta \vec Z}_{\alpha\beta}} \mbox{\Large\}}.
\label{eqn:atmod_sf}
\end{eqnarray}
Here, ${{\Delta \vec Z}}_{\alpha\beta} \equiv {\hat z} {\hat
z}\cdot(\vec \tau_\alpha - \vec \tau_\beta - \vec R)$, $\vec
u^{[1]}_{\alpha}$ is the first-order polarization vector and we refer
to the $e_\alpha$ as the ``atomic moduli''.

The atomic moduli as currently defined in Eq.~(\ref{eqn:atmod_sf})
provide a useful microscopic analysis of elastic response in bulk
systems which is similar in spirit to the decompositions used
previously in the study of bulk material
systems~\cite{vitek_c,jayanthi}.  To see that
Eq.~(\ref{eqn:atmod_sf}) indeed decomposes the overall elastic
response of bulk systems into atomic contributions coarse-grained over
distances on the order of the decay-length of the force-constant
matrix, we note first that for infinite bulk systems, elastic waves
are planar.  This implies that the first-order polarization vector
($\vec u^{[1]}_{\alpha}$) is uniform from primitive cell to primitive
cell and thus depends only on the local environment of each atom.
Next, we note that although the strain terms ($\Delta \vec
Z_{\alpha\beta}$) scale linearly with distance between atoms, the
terms which contribute to the final result are bounded in range by the
inter-atomic interactions ($\Phi(\vec R)$).  Thus, the atomic moduli
depend only on the local atomic environment over distances which the
decay of the force-constant matrix determines.

\begin{figure}
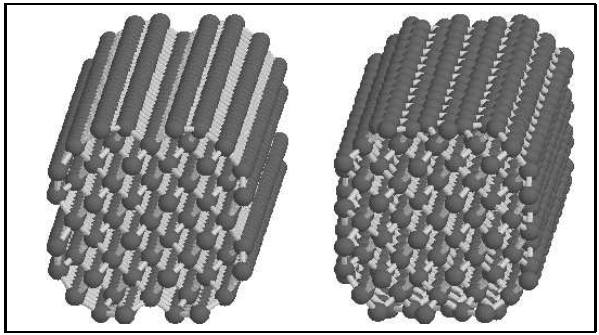

\begin{center}
\fbox{
\scalebox{0.425}{\ \ \ \ \ \ \includegraphics{fig1a.epsi}}\ \ \ \ \ \scalebox{0.425}{\includegraphics{fig1b.epsi}\ }
}
\end{center}
\caption{Atomic structure of silicon nanowires with an approximate
cross-sectional diameter of 1.5~nm.  The wires are viewed along the
longitudinal axis: $c(2\times 2)$ structure (left), $2\times 1$
structure (right).}
\label{fig:wires} 
\end{figure}

\section{Failure of straightforward approach in nanowires} \label{sec-fail}

We now demonstrate through direct calculations that the approach
outlined in the previous section gives unphysical results when applied
to nanoresonators.  Specifically, we study the behavior of
$[100]$-oriented nanoresonators of silicon, which recent {\em ab
initio} studies\cite{sismail} predict to undergo a size-dependent
structural phase transition between the two structures in
Figure~\ref{fig:wires} at a cross-section of $\sim$3~nm.  (The
interested reader may refer to Reference [21] for explicit details of
the microscopic structure of these wires.)  Initially, we work with
the Stillinger-Weber inter-atomic potential\cite{sw}, which suffices
for the exploration of general nanoelastic phenomena and which allows
study of cells with many thousands of atoms.  Later in the manuscript
(Section \ref{sec-cbames}) we use the Sawada tight-binding
model\cite{sawada} with modifications proposed by
Kohyama\cite{kohyama} to explore the correlation between our local
approach and the underlying electronic structure.  For all
calculations below, we fully relax the atomic coordinates, the
periodicity of the wire and, need be, the electronic structure.
Finally, we employ periodic boundary conditions along the
$z$-direction.

\begin{figure}
 \begin{center}
\rotatebox{90}{\hspace*{1in} $e_{\alpha}$ [eV]}
  \scalebox{0.4}[0.375]{\includegraphics{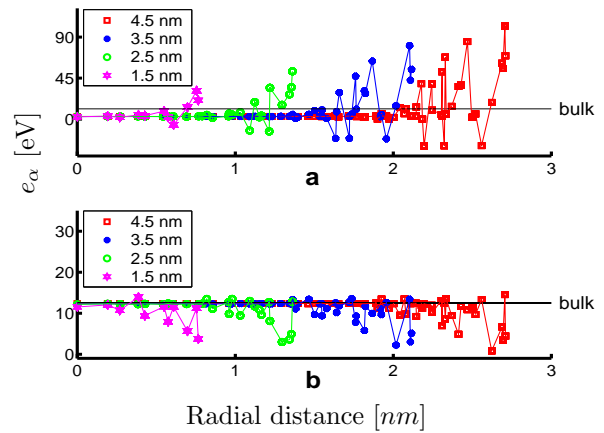}}
\\
Radial distance [$nm$]
 \end{center}
\caption{Predictions of atomic moduli $e_{\alpha}$ for $c(2\times 2)$
nanowires of varying diameter: (a) straightforward theory
(Eq.~(\ref{eqn:atmod_sf})) and (b) new theory
(Eq.~(\ref{eqn:atmod})).  The insets denote the approximate diameters
of the wires.  The value of the atomic modulus (Eq.~(\ref{eqn:atmod_sf})
or Eq.~(\ref{eqn:atmod})) is along the ordinate and radial distance of
the atom from the center line is along the abscissa.}
\label{fig:ecomp}
\end{figure}

Figure~\ref{fig:ecomp}a shows that the atomic moduli $e_\alpha$
predicted for nanowires using the straightforward approach of
Eq.~(\ref{eqn:atmod_sf}) are unphysical in that they depend upon the
macroscopic dimensions of the system and not simply on the local
environment of each atom.  In particular, the atomic
moduli on the surface grow linearly with the diameter of the wire and
the moduli in the center of the wire fail to approach the expected
bulk limit, $e_b \equiv Y_b/\rho_b$, where $Y_b$ is the Young's
modulus in bulk and $\rho_b$ is the number density in bulk.
(Note that these two effects are interelated, as the moduli must sum
to give the macroscopic value in the bulk limit.)

Although one has some freedom in choosing the terms used in the
perturbation expansion Eqs.~(\ref{eqn:sec})~and~(\ref{eqn:omegasq}),
for example to project Eq.~(\ref{eqn:sec}) against ${\bf u^{[0]}}$
instead of ${\bf u}$, all such expansions will lead to similar linear
scaling along the surface of the wire and approach an incorrect value
at the center of the wire.  Thus, straightforward application of the
method of long waves fails to result in a local, and hence physically
meaningful, description of elastic response in nanoscale systems, as
will straightforward variations thereon such as those in
References~[17]~and~[18].

\section{Method of long waves in nanowires} \label{sec-mlwn}

To cure the difficulties uncovered in the previous section, we proceed
by first identifying the cause of the pathological behavior and then
exploiting the freedom in Eq.~(\ref{eqn:atmod_sf}) to remove
this pathology.

The failure of the straightforward approach arises from the fact that
elastic waves in nanoresonators, or any system with free surfaces, are
not strictly planar.  In particular, the Poisson effect, which
the first order polarization vector ${\bf u}^{[1]}$ contains,
causes each atom to displace by an amount in direct proportion to its
distance from the center line of the system.  Eq.~(\ref{eqn:atmod_sf})
then leads directly to linear scaling of the atomic moduli at the
surfaces of the wire.

Defining an atomic elastic reponse dependent soley on the local enivornment
requires separation of extensive elastic effects from intensive
nanoscopic effects.  To seperate the extenive
motion in the first order polarization vector from that of its
intensive motion, we define the atomic displacements ($\vec
u^{rl}_{\alpha}$) as the intensive nanoscopic motions,
\begin{eqnarray} 
\vec u^{rl}_\alpha & \equiv & \vec u^{[1]}_{\alpha} - \left( -\sigma_x \hat
x \hat x \cdot - \sigma_y \hat y \hat y \cdot \right) \vec
\tau_\alpha, \label{eqn:decomp}
\end{eqnarray}
where $\sigma_{x,y}$ are the Poisson ratios.  There are three logical
choices for the Poisson ratios: some sort of local atomic definition,
an overall average for the wire, or the bulk values.  We choose to use
bulk Poisson ratios for a number of reasons.  First, a locally varying
definition makes it impossible to exploit the continuous rotational
and translational symmetries in the dynamical matrix, which we find
necessary to employ below in constructing atomic moduli with local
behavior.  Second, only by employing the bulk (rather than average)
Poisson ratios do we find a definition which approaches the
appropriate bulk value in the centers of wires of finite width.
Finally, we note that we always have the freedom of working with bulk
Poisson ratios because any motion along the surface in addition to
that resulting from bulk Poisson effect will not scale extensively
with the diameter of the wire and can therefore be incorporated into
the intensive atomic \mbox{displacements $\vec u^{rl}_{\alpha}$}.

After making the decomposition in Eq.~(\ref{eqn:decomp}), we next
employ the continuous rotational and translational symmetries of the
dynamical matrix to eliminate all extensive dependencies in
Eq.~(\ref{eqn:atmod_sf}).  Appendix~\ref{sec-app} outlines the
procedure for doing this, which then transforms
Eq.~(\ref{eqn:atmod_sf}) into
\begin{eqnarray}
e_\alpha & = & \sum_{\beta \vec{R}} \mbox{\Large\{} - {{\Delta \vec
r}_{\alpha\beta}} \cdot \Psi_{\alpha\beta}(\vec{R})\cdot {{\Delta \vec
r}_{\alpha\beta}} /2 + {\vec u^{rl}_\alpha} \cdot
\Phi_{\alpha\beta}(\vec{R})\cdot {\vec u^{rl}_\beta} \nonumber\\ && +
{{\Delta \vec r}_{\alpha\beta}} \cdot \Phi_{\alpha\beta}(\vec{R})\cdot
{\vec u^{rl}_\beta} - {\vec u^{rl}_\alpha} \cdot
\Phi_{\alpha\beta}(\vec{R})\cdot {{\Delta \vec r}_{\alpha\beta}}
\mbox{\Large\}}, \label{eqn:atmod}
\end{eqnarray}
where ${\Delta \vec r}_{\alpha\beta}$ represents the {\em total}
strain between atoms $\alpha$ and $\beta$,
$$
{\Delta \vec r}_{\alpha\beta} \equiv \left( -\sigma_x {\bf \hat x}
{\bf \hat x}- \sigma_y {\bf \hat y} {\bf \hat y} + {\bf \hat z} {\bf
\hat z} \right)\cdot(\vec \tau_\alpha - \vec \tau_\beta - \vec R),
$$
and $\Psi_{\alpha\beta}(\vec R)$ renormalizes as
$$
\Psi_{\alpha\beta}(\vec R) \equiv 2 \Phi_{\alpha\beta}(\vec R) -
\left( \text{Tr\,} \Phi_{\alpha\beta}(\vec R)
\cdot \Sigma
\right) \Sigma^{-1},
$$
where $\Sigma_{jk} \equiv \delta_{jk}
\sigma_j$, is a diagonal $3\times 3$ matrix with
elements $\sigma_x$, $\sigma_y$  and $\sigma_z \equiv -1$,
respectively and $\delta_{jk}$ is the Kronecker delta.

This new construction ensures that the modulus of each atom depends
only upon its local atomic environment because $\vec u^{rl}_{\alpha}$
no longer includes extensive motions and, although ${\Delta \vec
r}_{\alpha\beta}$ still depends on relative atomic distances, the
renormalized $\Psi$ decays as $\Phi$ does.  Thus, it is now the range
of the force-constant matrix which controls the size of the
neighborhood upon which each atomic modulus can depend.  Therefore,
the moduli of atoms in the interior now {\em must} correspond to the
expected bulk value, the moduli of the atoms on the surface now {\em
cannot} depend upon the extent of the system, and the resulting
description is physically meaningful.  Figure~\ref{fig:ecomp}b
illustrates the success of Eq.~(\ref{eqn:atmod}).

The fact that decomposition of elastic response into atomic level
contributions is not unique raises questions as to the physical
meaning of such a decomposition.  The new decomposition
Eq.~(\ref{eqn:atmod}) is the first which remains dependent only upon
local environment for systems with free surfaces.  Any other
definition {\em which respects locality} can only redistribute
portions of each atom's modulus among other atoms within a region of
extent comparable to the decay of the force-constant matrix.  Any sum
over such a region of the moduli will always be nearly the same.
Therefore, coarse grained over such regions, properly localized atomic
moduli become physically meaningful.  Section~\ref{sec-trans} and
Appendix~\ref{sec-appB} demonstrate this explicitly by comparing the
predictions for flexion from either properly localized moduli or
straightforwardly defined moduli, respectively.  To further
demonstrate that alternate local definitions are equivalent in this
coarse-grained sense, we have explored alternate local constructions.
In particular, while our present construction takes care to employ the
continuous symmetries of the dynamical matrix in such a way so as to
respect symmetry among the $x,y,z$ Cartesian coordinates, we have also
repeated the construction while treating the $x,y$ coordinates
symmetrically but not the $z$ coordinate and have found nearly
identical results for all of the applications below.

We close this section with a brief description of how the above
approach extends to any system of nanometer dimensions.  To be
considered small in this context, a dimension must be much smaller
than the wavelength of the distortions considered.  This manuscript
focuses on nanowires, systems with two small dimensions.  For a system
with one small dimension, for instance a plate with nanometer
thickness, the above approach develops in the same way with the one
minor change that the definition of $\vec u^{rl}_{\alpha}$
(Eq.~(\ref{eqn:decomp}) involves the Poisson effect only in the one
small dimension.  For an object with three small dimensions, for
instance adiabatic loading of a nano-object, the nature of the Poisson
effect depends upon the mode of loading, and the system should treated
as either of the above cases accordingly.  Finally, objects of no
small dimension, and hence no Poisson effect, can be described
straightforwardly as bulk-like using Eq.~(\ref{eqn:atmod_sf}).

\section{Transferability} \label{sec-trans}
To establish that our new atomic moduli are not mere convenient
mathematical constructions but are physically meaningful, we now
consider their transferability to phenomena not considered in their
original construction.  In particular, we consider flexion, where the
elastic distortion is no longer homogeneous throughout the
cross-section.

If our atomic moduli are indeed a measure of the local elastic
response, then under flexion the free energy per unit length will take
the form
\begin{eqnarray}
f&=&(1/L_c) \sum_\alpha e_\alpha u_\alpha^2/2, \label{eqn:free_at}
\end{eqnarray}
where $u_\alpha$ is a measure of the longitudinal strain which atom
$\alpha$ experiences.  For this form to be sensible, the diameter $D$
of the wire must not become comparable to the range of the
force-constant matrix so that the atoms which contribute to each
$e_\alpha$ all experience similar strains $u_\alpha$.  

Within continuum theory, uniform flexion with radius of curvature
$R$ corresponds to a longitudinal strain which varies linearly across
the wire, $u=x/R$.  (This holds to better than to two parts in
$10^3$ for all wires in our study.)  This would then predict a
flexural rigidity of
\begin{equation} 
F_{at} = \frac{1}{L_c} \sum_\alpha e_\alpha {x_\alpha}^2. \label{eqn:atmoF}
\end{equation}

\begin{figure}
 \centering 
  \scalebox{0.42}{\includegraphics{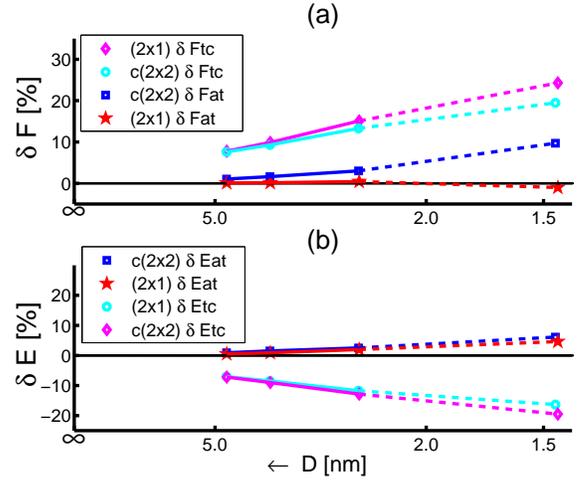}}
\caption{Fractional error as a function of inverse diameter $1/D$ in
predicting (a) flexural rigidity from extensional properties ($\delta
F_{at}$, Eq.~(\ref{eqn:datmoF}) and $\delta F_{tc}$,
Eq.(\ref{eqn:dcntF})) and (b) extensional rigidity from flexural
properties ($\delta E_{at}$, Eq.~(\ref{eqn:datE}) and $\delta E_{tc}$,
Eq.~(\ref{eqn:dtcE})).  Note, for convenience, the abscissa is labeled
by $D$ and not $1/D$.}
\label{fig:EandF}
\end{figure}

Figure~\ref{fig:EandF}a shows the fractional error ($\delta F_{at}$)
\begin{eqnarray}
\delta F_{at} \equiv (F_{at} - F)/F \label{eqn:datmoF},
\end{eqnarray}
in predicting the flexural rigidity from Eq.~(\ref{eqn:atmoF}), where
$F$ is determined directly through numerical calculations.  The figure
shows that these errors are indeed quite small.  Note that use of the
straightforward definition in Eq.~(\ref{eqn:atmod_sf}) with its
unusually scaling surface moduli leads to invalid
predictions for flexion.  (Appendix~\ref{sec-appB} shows this directly
through scaling arguments.)  As a result, to have {\em predictive}
power, definitions of atomic moduli must properly account for the
Poisson effect.

To demonstrate that our new approach has greater predictive power than
traditional continuum approaches, the figure also shows the
fractional error ($\delta F_{tc}$) in predicting the flexural rigidity
when using the traditional continuum relation
\begin{eqnarray}
F_{tc} &=& E(I/A), \label{eqn:cntF} \\ 
\delta F_{tc} &=& (F_{tc} - F)/F \label{eqn:dcntF}
\end{eqnarray}
where $I/A$ defines the mean bending moment, which we define
unambiguously as
\begin{eqnarray}
I/A &=& (1/N_c)\sum_{\alpha}x_{\alpha}^{2}. \label{eqn:mmoi}
\end{eqnarray}

The linear behavior of the fractional error $\delta F_{tc}$ as a
function of $1/D$ indicates that continuum theory does not properly
account for surface effects, a result of the fact that flexion places
a larger emphasis on the surface than does extension.  The dramatic
improvement from the use of Eq.~(\ref{eqn:atmoF}) arises because the
atomic moduli place proper emphasis on the surface and on the
interior, as they properly treat each atomic environment locally.  The
atomic moduli therefore properly account for elastic fluctuations
along the cross-section of the wire which are on scales too small for
traditional continuum theories to capture.

Finally, we note that the only appreciable error within the new
framework occurs for the smallest wires (D$\approx 1.5$nm).  At this
point, the cross-sectional dimension becomes comparable to the range
of the force-constant matrix, and Eq.~(\ref{eqn:free_at}) represents
an improper use of the physical concept of atomic moduli.  As
described above (second from last paragraph in
Section~\ref{sec-mlwn}), only sums of atomic moduli over regions of
extent comparable to the range of the force-constant matrix carry
physical meaning.  Any sum sensitive to variations over shorter
scales, as is Eq.~(\ref{eqn:free_at}) when limited to wires narrower
than the range of the force-constant matrix, cannot be depended upon
to lead to meaningful results.  This underscores the fact that
properly construed atomic moduli are not truly atomic-level quantities
but a concept coarse grained over the range of the force constant
matrix.  As we have seen, however, this coarse-graining is on scales
significantly smaller than those captured by traditional continuum
theory.

\section{Extension from flexion} \label{sec-fte}
We now apply the above defined atomic moduli to derive a new relation
for the extensional modulus in terms of the flexural modulus and other
experimentally accessible observables, a relation often needed in
experimental analyses~\cite{wong,treacy,salvetat,osakabe}.  We then
show that the new relation is much more accurate than the standard
continuum-theory based relation currently employed in experimental
analyses.  Finally, we employ our concept of atomic moduli to provide
quantitative insight into the improvement of our new relation over the
traditional continuum relation.

The basis for the following analysis is the fact that
Eq.~(\ref{eqn:atmoF}) gives a very good estimate of the true flexural
modulus, as Figure~\ref{fig:EandF}a confirms.  Using the exact
relation for $E$, Eqs.~(\ref{eqn:thermofree})~and~(\ref{eqn:atmod}),
and Eq.~(\ref{eqn:atmoF}), one derives the following leading-order
prediction $E_{at}$ for the extensional modulus,
\begin{eqnarray}
E_{at} & = & \frac{1}{2}\left( \frac{F}{I/A} +
Y_b\frac{\lambda}{\rho_b} \right),\label{eqn:atEfromF}
\end{eqnarray}
with a predicted error $\delta E_{at}^{(p)}$ of
\begin{eqnarray}
\delta E_{at}^{(p)}  &=&  \frac{N_s}{E L_c} \left\{ B_{at}\left[\langle e_s
\rangle_{x^{2}} - e_b\right]  + \left[\langle e_s \rangle_{x^{2}} - \langle e_s
\rangle\right] \right\}. \label{eqn:error}
\end{eqnarray}
Here, $N_s$ is the number of ``surface'' atoms, defined as those for
which $e_\alpha$ differs significantly from the limiting bulk value
$e_b$, $\langle e_s\rangle_{x^2}$ is the inertia weighted average
surface moduli
$$
\langle e_s\rangle_{x^{2}} \equiv \frac{\sum_s e_s x_{s}^{2}}{\sum_s
x_{s}^{2}},
$$
with the sums ($\sum_s$) ranging over ``surface'' atoms,  $\langle
e_s\rangle$ is the average surface moduli
$$\langle e_s\rangle \equiv \frac{1}{N_s}\sum_s e_s
$$ 
and
$$
B_{at}\equiv(1/2) \frac{\frac{1}{N_s}\sum_s x_{s}^{2}}{\frac{1}{N_c}\sum_c x_{c}^{2}} -1,
$$
where $\sum_c$ implies sums over all atoms in the unit cell.   This
result holds for any division of the atoms into ``surface'' and
``bulk'' to the extent that each ``bulk'' atom has atomic modulus
$e_b$.

Figure~\ref{fig:EandF}b shows the relative error $\delta E_{at}$,
\begin{eqnarray}
\delta E_{at} = (E_{at} - E)/E, && \label{eqn:datE}
\end{eqnarray}
between the extensional modulus $E$ determined directly from numerical
calculation and as determined from our new relation,
Eq.~(\ref{eqn:atEfromF}).  (Note that $\delta E_{at}^{(p)} = \delta
E_{at}$ exactly when the moduli prediction $F_{at}=F$ holds.)  As the
relevant point of comparison, the figure also shows the relative error
$\delta E_{tc}$,
\begin{eqnarray}
\delta E_{tc} = (E_{tc} - E)/E, \label{eqn:dtcE}
\end{eqnarray}
associated with the traditional continuum result
\begin{equation} \label{eqn:cntE}
E_{tc} = F \frac{A}{I}.
\end{equation}
Hence, Eq.~(\ref{eqn:atEfromF}) is much more accurate than the
standard continuum result Eq.~(\ref{eqn:cntE}).

To understand the improvement of the new relation,
Eqs.~(\ref{eqn:thermofree})~and~(\ref{eqn:atmoF}) may also be combined
to yield a prediction for the fractional error in the traditional
continuum analysis,
\begin{equation}
\delta E_{tc}^{(p)} =  \frac{N_s}{E L_c} \left\{
B_{tc}\left[\langle e_s \rangle_{x^{2}} - e_b\right] + \left[\langle
e_s \rangle_{x^{2}} - \langle e_s \rangle\right] \right\},
\label{eqn:errorct}
\end{equation}
which takes {\em precisely} the same form as Eq.~(\ref{eqn:error})
except for the change in the prefactor in the first term from $B_{at}$
to $B_{tc} \equiv 2B_{at} + 1$.  For continuous wires of homogeneous
circular or regular polygonal cross-section, we have exactly $B_{at} =
0$.  Thus, generally we expect $B_{at}$ to be close to zero and
$B_{tc}$ to be close to unity.  We now note that the term in the first
set of square brackets ($\left[\langle e_s \rangle_{x^{2}} -
e_b\right]$) in both Eq.~(\ref{eqn:error}) and Eq.~(\ref{eqn:errorct})
is an average difference between surface and bulk atoms and,
therefore, is generally much larger than the term in the second set of
square brackets $\left[\langle e_s \rangle_{x^{2}} - \langle e_s
\rangle\right]$, which is the difference between two differently
weighted averages over the surface atoms.  Thus the larger term nearly
vanishes in our new relation, Eq.~(\ref{eqn:error}), but not in the
traditional continuum relation, Eq.~(\ref{eqn:errorct}).  From this
analysis, we see that the reason why the traditional continuum
relation has larger errors is that it does not properly differentiate
between the local surface and bulk environments.

The atomic moduli also lead to a quick, intuitive argument to
understand the improvement of Eq.~(\ref{eqn:atEfromF}) over
Eq.~(\ref{eqn:cntE}).  From our results we know that the surface
moduli can be quite different than those of the bulk.  It is also known
that flexion places larger emphasis on the surface than does
extension.  If the average surface modulus is less/more than that of
the bulk then the first term in Eq.~(\ref{eqn:atEfromF}) ($F \cdot
A/I$) will underestimate/overestimate the extensional rigidity, while
the second term will overestimate/underestimate it.  Therefore, errors
will tend to cancel in the average of the two.

\section{Correlation between local atomic moduli and electronic
structure} \label{sec-cbames}

We now explore the local atomic moduli as a link between local elastic
properties and the underlying electronic structure.  To do this, we
have calculated the atomic moduli, Eq.~(\ref{eqn:atmod}), for both the
$c(2\times 2)$ and $(2\times 1)$ structures, using the
Sawada~\cite{sawada} tight-binding model with modifications proposed
by Kohyama\cite{kohyama}.  We have studied various wires, all of which
give similar results.  For brevity, we here only report on wires with
cross-sectional diameter $\approx 2.4$nm.  The supercell of our
calculation is four bulk cubic lattice constants long in the periodic
direction, and hence sampling the Brillouin zone at the $\Gamma$ point
in the electronic structure calculations is more than sufficient.

\begin{figure}
 \centering
 \parbox{1.75in}{\scalebox{0.4}{\includegraphics{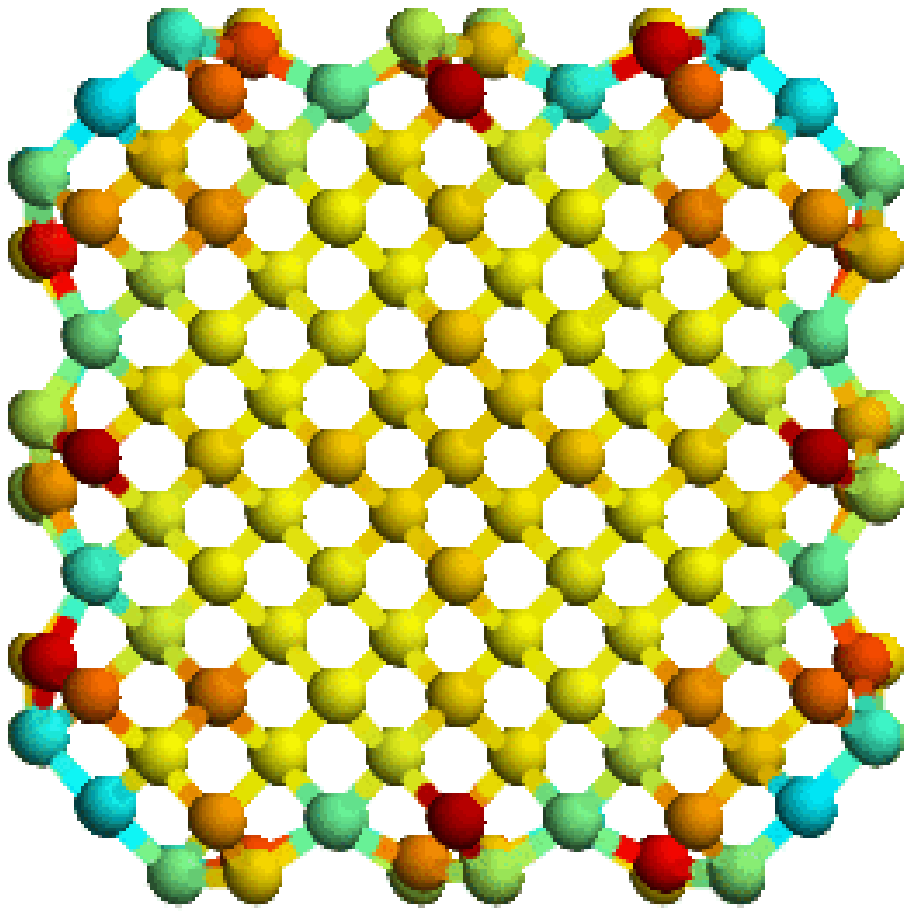}}}
 \parbox{1.25in}{\scalebox{0.2}{\includegraphics{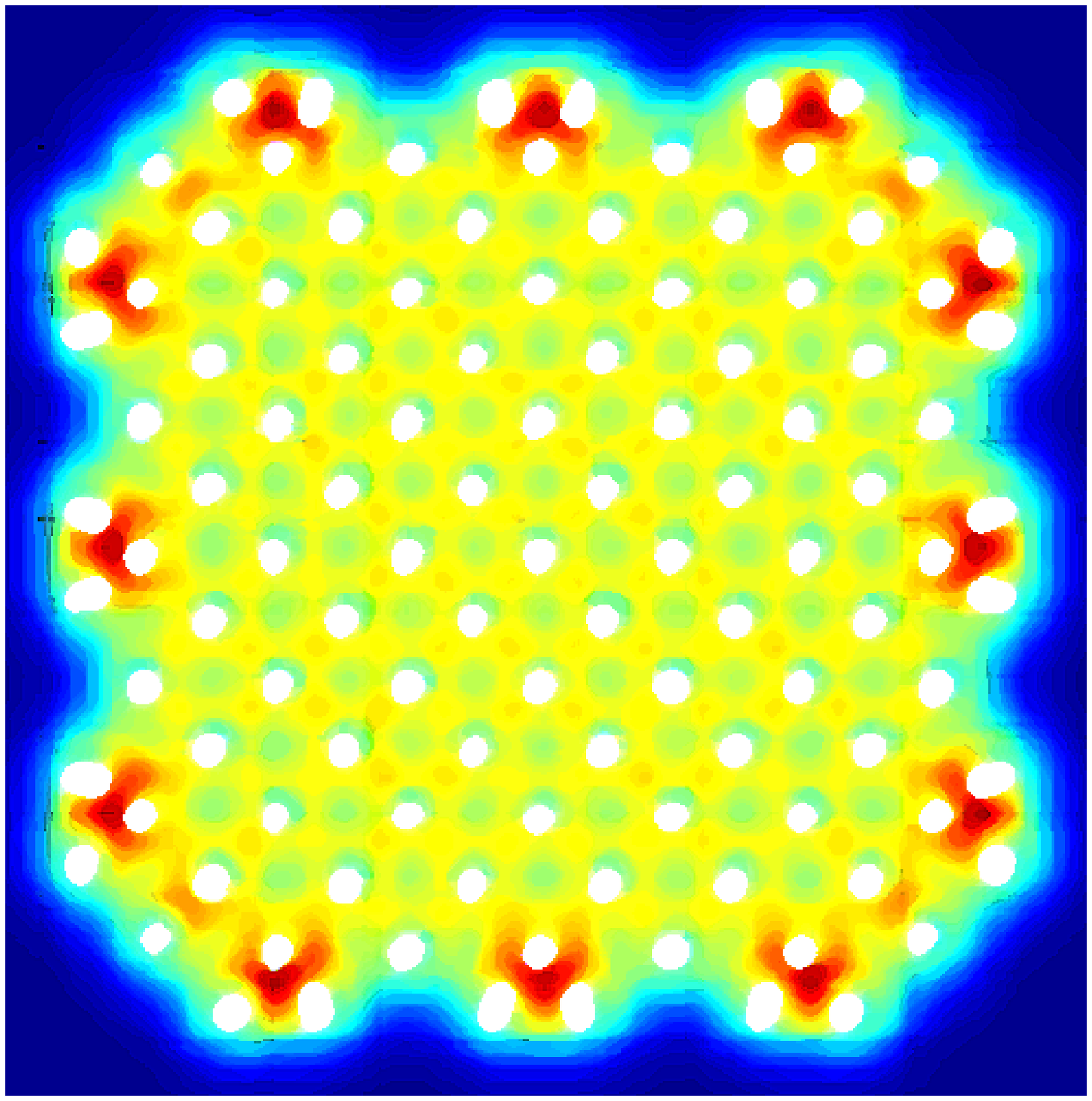}}}
\caption{\ctbt wire, with cross-section $\approx 2.4$nm.  Results are
from the tight-binding model described in the text.  Left: atomic
moduli color coded from small to large: blue-green-yellow-red.  Right:
valence electronic density projected (integrated along the wire axis)
onto the cross-section color coded from low to high:
blue-green-yellow-red.  White dots in the right figure indicate the
location of atom cores.  Note the correlation of large atomic moduli
with large charge density along the surface.}
\label{fig:tbc2x2}
\end{figure}

\begin{figure}
 \centering 
 \parbox{1.75in}{\scalebox{0.4}{\includegraphics{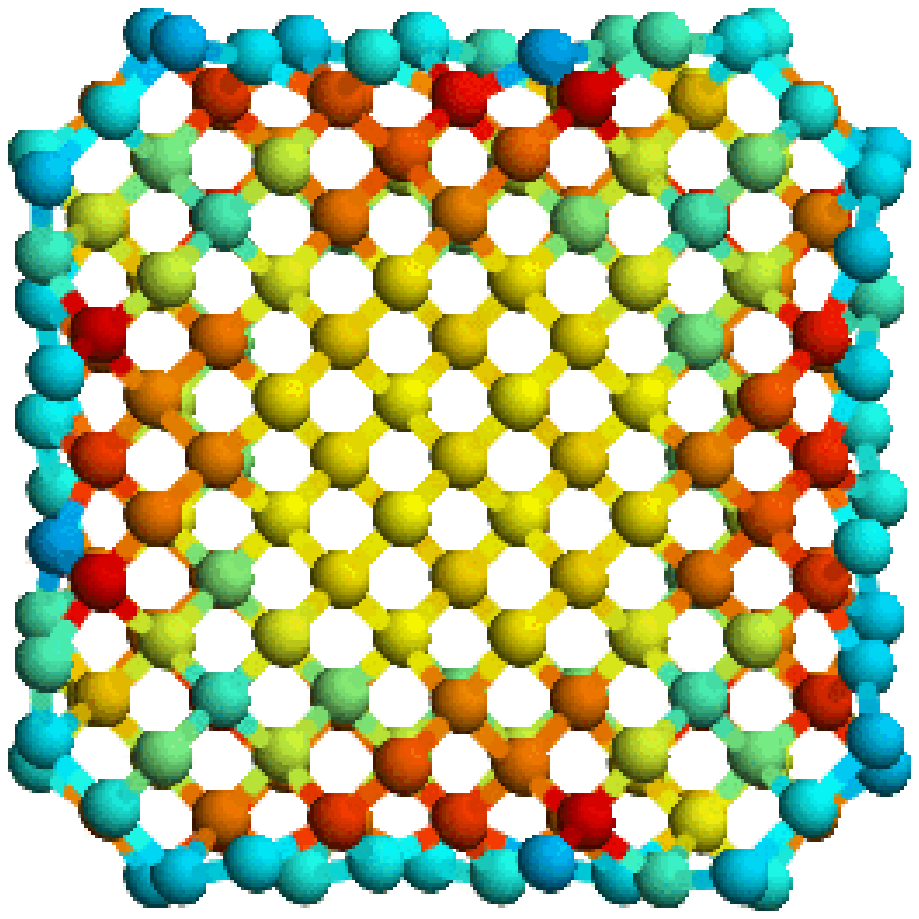}}} \parbox{1.25in}{\scalebox{0.2}{\includegraphics{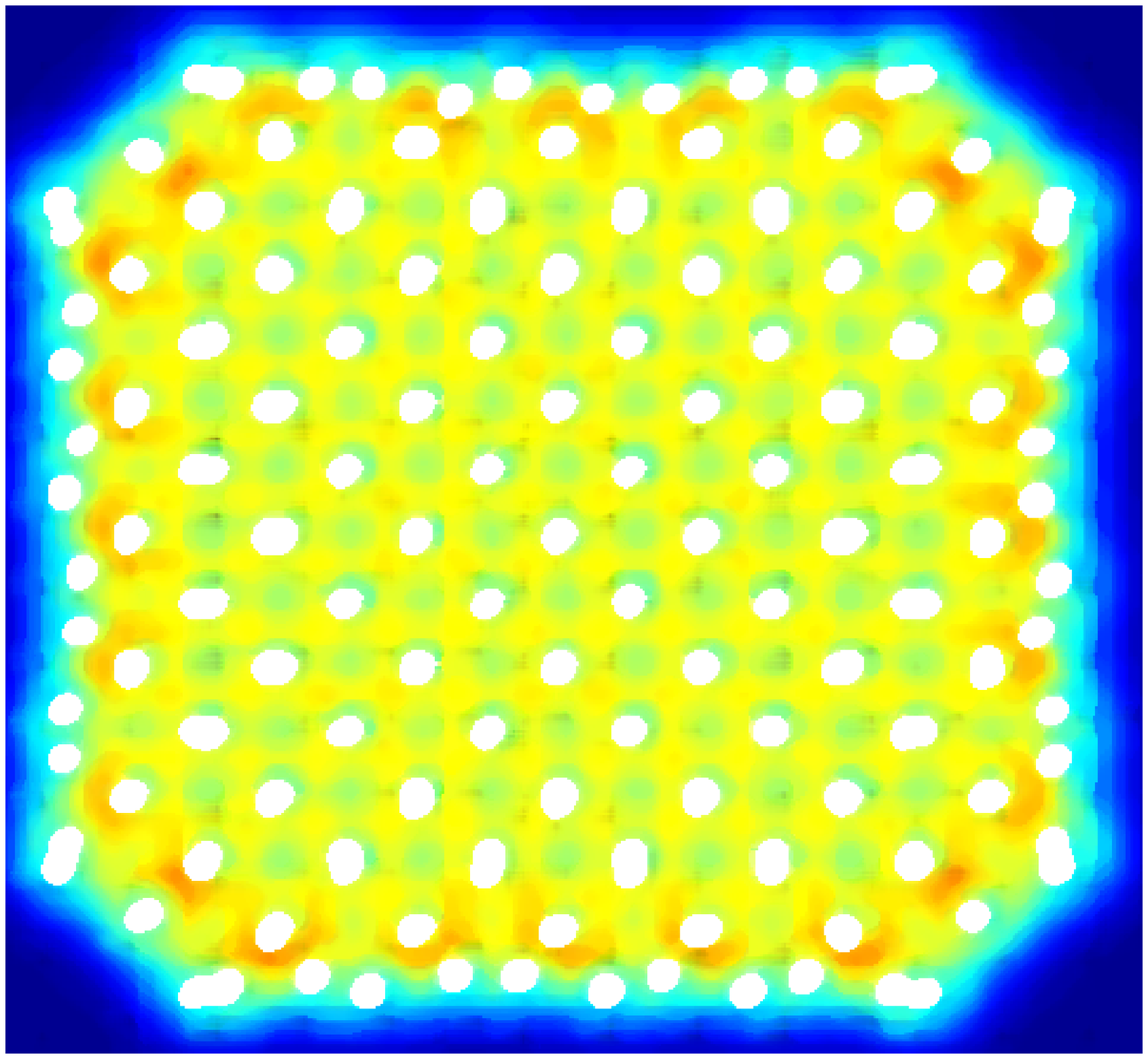}}}
\caption{\tbo wire, with cross-section $\approx 2.4$nm: same
conventions as Figure~\protect{\ref{fig:tbc2x2}}.}
\label{fig:tb2x1}
\end{figure}

The left panels of Figures~\ref{fig:tbc2x2}~and~\ref{fig:tb2x1} show
the resulting atomic moduli, with values color coded so that yellow
corresponds to the bulk value (17.1 eV/atom).  The figure shows that
deviations from this value concentrate near the surface in patterns
characteristic of the structure of the wire.  Moduli near the surface
fluctuate widely, ranging from 6-27 eV/atom for the $c(2\times 2)$
structure and from 4-26 eV/atom for the $2\times 1$ structure.

To allow comparison with the underlying electronic structure, the
right panels in the figures display the valence charge densities from
the tight-binding calculation projected (integrated along the wire
axis) onto the cross-section of the wire.  (To compute the electron
density from the tight-binding coefficients, we employed orbitals from
a density functional calculation of the silicon atom.)  The figures
display the electron densities using a color map similar to that
employed for the atomic moduli.  Intriguingly, there is an apparent
correlation between the values of the atomic moduli and the underlying
electron density.  In particular, large/small atomic moduli correlate
with regions of large/small electron density in
Figure~\ref{fig:tbc2x2}, indicating that charge distribution along the
surface of these wires greatly affects the local elastic properties
and thereby the overall elastic response, particularly to flexion
which emphasizes surface effects.  Figure~\ref{fig:tb2x1} exhibits a
similar correlation, but not as pronounced.  Note that, in this figure,
coincidence of red atomic moduli just under the surfaces of the first
layer of atoms correlates with red charge densities in the same
location.

Unlike in the classical potential case, where properly defined surface
moduli are systematically lower or equal to the bulk value due to
decrease in the number of bonds (Figure~\ref{fig:ecomp}b), we find
that in a quantum model, surface moduli may even greatly exceed the
bulk value due to changes in local charge density which can enhance
the mechanical strength of bonds
(Figures~\ref{fig:tbc2x2}~and~\ref{fig:tb2x1}.)  This contrast
underscores both the importance of considering contributions of the
electronic structure to mechanical response and the need for a
definition of atomic moduli which can be computed from physical
observables obtainable from electronic structure calculations.

The ultimate use of atomic level moduli is to understand mechanical
response.  Table~\ref{table:tab1} compares the errors from both
continuum theory and the use of atomic-level moduli in predicting
mechanical response for the two wires under consideration in this
section.  The quantities compared exactly parallel those of the
previous section.  The first two columns of the table consider
prediction of flexural response from continuum theory and our atomic
moduli, respectively, and the second two columns consider prediction
of extensional response from flexural response using either the
traditional continuum relation or our new relation, respectively.  The
table shows that the new relations,
Eqs.~(\ref{eqn:atmoF})~and~(\ref{eqn:atEfromF}) are again very
accurate.  The table also shows that these predictions are superior to
the corresponding continuum results,
Eqs.~(\ref{eqn:cntF})~and~(\ref{eqn:cntE}), respectively.

\begin{table}
\centering
\begin{tabular}{|c|c|c|c|c|c|}
\hline Structure & $\delta F_{tc}$ & $\delta F_{at}$ & & $\delta
E_{tc}$ & $\delta E_{at}$ \\ \hline \tbo & 13.1\% & 2.86\% & & -11.6\%
& 1.78\%\\ \hline \ctbt & 4.91\% & 1.79\% & & -4.68\% & 2.02\% \\
\hline
\end{tabular}
\caption{Comparison of errors between the traditional continuum theory
and the atomic moduli description when predicting flexural response
through Eqs~(\ref{eqn:cntF})~and~(\ref{eqn:atmoF}) (first and second
columns, respectively) and when inferring extensional response from
flexural response through
Eqs.~(\ref{eqn:cntE})~and~(\ref{eqn:atEfromF}) (third and fourth
columns, respectively).}
\label{table:tab1}
\end{table} 

Interestingly, the continuum predictions for the $c(2\times 2)$ wire
are fairly reliable.  The atomic moduli provide an avenue for
understanding this as well.  Fluctuations in the moduli in the
$c(2\times 2)$ wires are localized and hence average out over regions
of extent comparable to the decay of the force-constant matrix.
Moreover, in this particular case they tend to average to values close
to that expected of the bulk.  Without meaningful fluctuations on the
length scales of the decay of the force-constant matrix, we expect
continuum theory to perform well for this wire.  In contrast, the $2
\times 1$ wire exhibits much more systematic variations in the moduli.
The outermost surface atoms have a consistent and significantly
reduced modulus, and there is also a clear significant and systematic
variation in the moduli throughout the cross-section of the wire.
Because this second wire does exhibit meaningful fluctuation over
distances comparable to the decay range of the force constant matrix,
we expect traditional continuum relations to give particularly poor
results, underscoring the importance of the local atomic-level moduli
description.

\section{Conclusions} 
This manuscript presents the first definition of atomic-level elastic
moduli for nanoscale systems which are defined in terms of physical
observables, correctly sum to give the exact overall elastic response
and depend only on the local environment of each atom.  Although these
moduli are not necessarily uniquely defined, their sum over regions of
extent comparable to the range of the force constant matrix is
physically meaningful and may be used to make accurate predictions of
mechanical response.  The moduli resulting from our formulation
transfer to different modes of strain and correctly account for
elastic fluctuations on the nanoscale.  They also lead to a
quantitative understanding of when traditional continuum relations
breakdown and how to properly correct them properly.  Specifically, we
demonstrated a more accurate method for relating extensional and
flexural properties.  These moduli provide a clear and natural method
for distinguishing {\em mechanically} between ``surface'' atoms and
``bulk'' atoms and give insight into the correlation between the local
mechanical response and the underlying electronic structure.  Finally,
these moduli allow the identification of which atomic arrangements
lead to more pliant or stiffer response opening the possibility of
their use as a tool to aid in the rational design of nanostructures
with specific mechanical properties.

\section{Acknowledgments}
D.E.~Segall acknowledges support of his Department of Physics
fellowship and the department's continuing support.  S.~Ismail-Beigi
acknowledges support of the MIT MRSEC Program of the National Science
Foundation under award number DMR 94--00334.  Computational support
was provided by the MIT Xolas prototype SUN cluster and the Cornell
Center for Materials Research Shared Experimental Facilities,
supported through the NSF MRSEC Program DMR--9632275.

\appendix

\section{Use of continuous symmetries to produce local moduli} \label{sec-app}

This appendix outlines the use of rotational and translation
symmetries of the force constant matrix to reformulate the ill-defined
atomic moduli in Eq.~(\ref{eqn:atmod_sf}) into the well-defined form
in Eq.~(\ref{eqn:atmod}).
 
From continuous translational symmetry, all force-constant matrices
obey
\begin{eqnarray}
\sum_{\beta,\vec R} \Phi_{\alpha,\beta}(\vec R)\cdot \vec c_{\beta}
&=& 0, \label{eqn:tran}
\end{eqnarray}
for any vector $\vec c_{\beta} = \vec c$ that is the constant vector
for all atoms in the unit cell.  Moreover, the
$SO(3)$ rotational symmetries imply
\begin{eqnarray}
\sum_{\beta,\vec R} \Phi_{\alpha q,\beta s}(\vec R) r_{\beta t} =
\sum_{\beta,\vec R} \Phi_{\alpha q,\beta t}(\vec R) r_{\beta s},
\label{eqn:rot}
\end{eqnarray}
where the $q,s,t$ correspond to one of the $x,y,z$ Cartesian
coordinates, $\Phi_{\alpha q,\beta s}(\vec R)$ is the $q,s$ component
of $\Phi_{\alpha,\beta}(\vec R)$ and $r_{\beta s}$ corresponds to the
$s$ component of the position vector $\vec r_{\beta}$.

The ill-defined atomic moduli,
Eqs.~(\ref{eqn:thermofree})-(\ref{eqn:atmod_sf}), contain divergent
terms which are in one of the following two forms:
\begin{eqnarray}
 && \sum_{\alpha,\beta  \vec  R}  \sigma_{s}  r_{\alpha  s}\Phi_{\alpha
s,\beta  t}(\vec R)u^{rl}_{\beta  t},  \label{eqn:firstterm}  
\end{eqnarray}
or
\begin{eqnarray}
 && \sum_{\alpha,\beta \vec  R} \sigma_s r_{\alpha  s}\Phi_{\alpha s,\beta
t}(\vec R)r_{\beta t}\sigma_t. \label{eqn:secondterm}
\end{eqnarray}
All of these terms scale linearly along the surface of the wire and
give rise to the linear scaling of the surface moduli with system size
evident in Figure~\ref{fig:ecomp}a.

From Eq.~(\ref{eqn:tran}), one can set Eq.~(\ref{eqn:firstterm}) equal
to
\begin{equation}
 \sum_{\alpha,\beta \vec R}
\sigma_{s} (r_{\alpha s} - r_{\beta s})\Phi_{\alpha s,\beta t}(\vec
R)u^{rl}_{\beta t}.
\label{eqn:firstnew}
\end{equation}
Using both Eqs.~(\ref{eqn:tran})~and~(\ref{eqn:rot}), one can set
Eq.~(\ref{eqn:secondterm}) equal to
\begin{eqnarray}
-\sum_{\alpha,\beta \vec R}& \sigma_s\left[\right.(r_{\alpha s} -
 r_{\beta s})\Phi_{\alpha s,\beta t}(\vec R)(r_{\alpha t} - r_{\beta
 t}) & \nonumber \\ & - \frac{1}{4} (r_{\alpha s} - r_{\beta
 s})\Phi_{\alpha t,\beta t}(\vec R)(r_{\alpha s} - r_{\beta s}) &
 \nonumber \\ & - \frac{1}{4} (r_{\alpha t} - r_{\beta t})\Phi_{\alpha
 s,\beta s}(\vec R)(r_{\alpha t} - r_{\beta t}) &
 \left.\right]\sigma_t. \label{equ:secondnew}
\end{eqnarray}

The above terms now only depend on relative distances over a range
controlled by the decay of the force-constant matrix and therefore no
longer scale linearly with the size of the system.  Combining the
above transformations with the separation of the extensive motion from
the intensive motion in the first-order polarization vector,
Eq.~(\ref{eqn:decomp}), then results in the well defined form
Eq.~(\ref{eqn:atmod}).

\section{Failure of straightforward approches to predict flexion} \label{sec-appB}

We now demonstrate that although the moduli defined in
Eq.~(\ref{eqn:atmod_sf}) indeed sum to give the correct overall
extensional response, their unphysical scaling properties lead to
invalid physical predictions when used in other contexts and,
therefore, that they are ill-defined.  In particular, we use simple
scaling arguments to prove that these ill-defined moduli give a
nonnegligible error in predicting the flexural rigidity in the
continuum limit.

Consider a circular wire whose radius $R$ is sufficiently large such
that continuum theory applies.  From Figure~\ref{fig:ecomp}a we see
that the surface moduli defined by Eq.~(\ref{eqn:atmod_sf}) scale
linearly with $R$.  Because of this, and in order to arrive at an
analytic result for the error in predicting the flexural rigidity, we
can assume, to a good approximation, that all surface moduli $e_s$
are proportional to their radial distance $r$ from the center of the
wire,
\begin{equation}
e_s = \bar e_s r. \nonumber
\end{equation}
Here, $\bar e_s$ is the same for all surface atoms.  Next, we define
$R_i$ as the inner radius such that all atoms at position $r < R_i$
are in the bulk and all atoms with positions $r > R_i$ are on the
surface. Finally, we define $\Delta R \equiv R-R_i$.

Because of the facts that the surface moduli scale linearly with
the system and that the sum of all atomic moduli must equal the
extensional rigidity, we conclude that the moduli in the bulk region of the
wire, derived from Eq.~(\ref{eqn:atmod_sf}), $e_b'$ cannot equal the
average value of the bulk material $e_b$.  (Figure~\ref{fig:ecomp}a
also evidences this behavior.)  The two above facts imply that the
following equality must hold,
\begin{eqnarray}
\int_{R_i}^{R} e_s r\,dr + \int_0^{R_i} e_b' r\,dr = \int_0^R e_b r\,dr, &&
\nonumber
\end{eqnarray}
or to leading order in $\Delta R/R_i$,
\begin{eqnarray}
\bar e_s \Delta R(1 + \Delta R/R_i)  + e_b'/2 = e_b (1/2 + \Delta R/R_i).
\label{eqn:ext}
\end{eqnarray}

In the continuum limit, the traditional continuum relation
Eq.~(\ref{eqn:cntF}) holds, and therefore the fractional error in
predicting the flexural rigidity from the ill-defined atomic moduli
Eq.~(\ref{eqn:atmod_sf}) is equal to
\begin{eqnarray}
  \delta F_{ill-at} &  = & \frac{\int_{0}^{2\pi}d\theta
\, \left( \int_{R_i}^{R} dr \, x^2 e_s r + \int_0^{R_i} dr \, x^2 e_b'
r \right) }{\int_{0}^{2\pi}d\theta \, \int_0^R dr\, x^2 e_b r} - 1
\nonumber
\end{eqnarray}
or to leading order in $\Delta R/R_i$, 
\begin{eqnarray}
\delta F_{ill-at} & = & \frac{\bar e_s \Delta R(1 + 2\Delta
R/R_i) + e_b'/4 - e_b'\Delta R/R_i}{e_b/4} - 1.
\label{eqn:flx}
\end{eqnarray}
Using Eq.~(\ref{eqn:ext}) to solve for $\bar e_s \Delta R$,
Eq.~(\ref{eqn:flx}) becomes
\begin{eqnarray}
\delta F_{ill-at} & = & 1 - \frac{e_b'}{e_b} +
\frac{2(e_b-e_b')}{e_b}\Delta R/R. \nonumber
\end{eqnarray}
Therefore, in the continuum limit, the ill-defined moduli do not
approach the correct result and thus give a prediction which is even
worse than that of traditional continuum theory.

\end{document}